# Ultrafast Ultraviolet C and Visible Laser Light with Broadly Tunable Spectrum

**Dimitar Popmintchev,**[1,3] **Aref Imani,**[1] **Paolo Carpegiani,**[1] **Joris Roman,**[1] **Siyang Wang,**[2] **Jieyu Yan,**[2] **Sirius Song,**[2] **Ryan Clairmont,**[2] **Ayush Singh,**[2] **Edgar Kaksis,**[1] **Tobias Flöry,**[1] **Audrius Pugžlys,**[1] **Andrius Baltuška,**[1] **Tenio Popmintchev**[1,2,4]

[1]*Photonics Institute, TU Wien, Vienna A-1040, Austria*
[2]*University of California San Diego, Physics Department, Center for Advanced Nanoscience, La Jolla, CA 92093, USA*
[3]dimitar.popmintchev@tuwien.ac.at
[4]tenio.popmintchev@physics.ucsd.edu

**We demonstrate a versatile technique for generating continuously wavelength-tunable laser waveforms, with mJ pulse energies and ultrashort pulse durations down to few-cycle in the ultraviolet C and visible spectral ranges. Using the processes of self-phase modulation or Raman-induced spectral broadening, we substantially expand the spectrum of a femtosecond $1030\ nm\ Yb{:}CaF_2$ laser, allowing for an extensive wavelength-tunability of the second and fourth harmonics of the laser within the visible and ultraviolet C spectral regions at $460 - 580\ nm$ and $230 - 290\ nm$. In addition, our approach exploits nonlinearly assisted self-compression in the second harmonic upconversion of the spectrally broadened infrared pulses with high-order dispersion. This results in $8 - 30$ femtosecond pulses in the visible with an intensity enhancement of up to 30 times. Such an ultrashort visible and ultraviolet C source is ideal for investigating ultrafast dynamics in molecules, solids, and bio and nano systems. Furthermore, this tunable light source enables the generation of bright, narrow-bandwidth, continuously wavelength-tunable, coherent light in the extreme ultraviolet to soft X-ray spectral region. Theoretically, the X-ray pulse structure can consist of sub-200 $as$ pulse trains, making such a source highly suitable for dynamic multidimensional imaging. This includes coherent diffractive imaging of ferromagnetic nanostructures where resonant X-ray scattering is essential, as well as X-ray absorption spectroscopies of advanced quantum materials at pico-nanometer spatial and atto-femtosecond temporal scales.**

**Introduction.** The development of intense, ultrashort laser pulses has revolutionized and advanced various optical fields, such as medical imaging, optical metrology, high-precision spectroscopy, laser-matter interactions, frequency conversion towards the X-ray or THz spectral regions, control and study of chemical reactions, optical telecommunications, medical surgeries, and more. These sources have not only enabled probing or manipulating dynamical processes at the ultrafast pico-to-zeptosecond temporal and micro-to-picometer spatial scales but also have opened up new fields. In particular, in the past decades, they have also enabled the field of attosecond and coherent EUV and X-ray science through the extremely nonlinear process of high-order harmonic generation. The tabletop EUV – soft X-ray sources have reached a new level of performance, delivering fully coherent light with high brightness, unprecedented spatial coherence, and attosecond-scale temporal structure. [1-7]. Coherent X-rays have the ability to penetrate thick, opaque objects, visualize subwavelength-nanostructures, control spin and magnetic currents, and access materials' chemistry and structure at picometer-length and femto-zeptosecond temporal scales, among other applications. [7-12]. However, wavelength-tunability has been a major limitation thus far for these sources. This feature is essential for identifying materials and their physical and chemical properties through the absorption-edge fingerprints. Hence, a high brightness, narrow-bandwidth, wavelength-tunable, EUV – soft X-ray source with an attosecond pulse structure and a characteristically low attosecond chirp would have been the ultimate source for ultrafast coherent magnetic imaging or X-ray absorption spectroscopies. Furthermore, such a source is of significance for water window (280-530eV) implementation, where bio-building structural blocks absorb more than water, enabling in vivo optimization and study of biological or high-tech specimens for applications such as gain-of-function surveys, data storage, energy storage, and more.

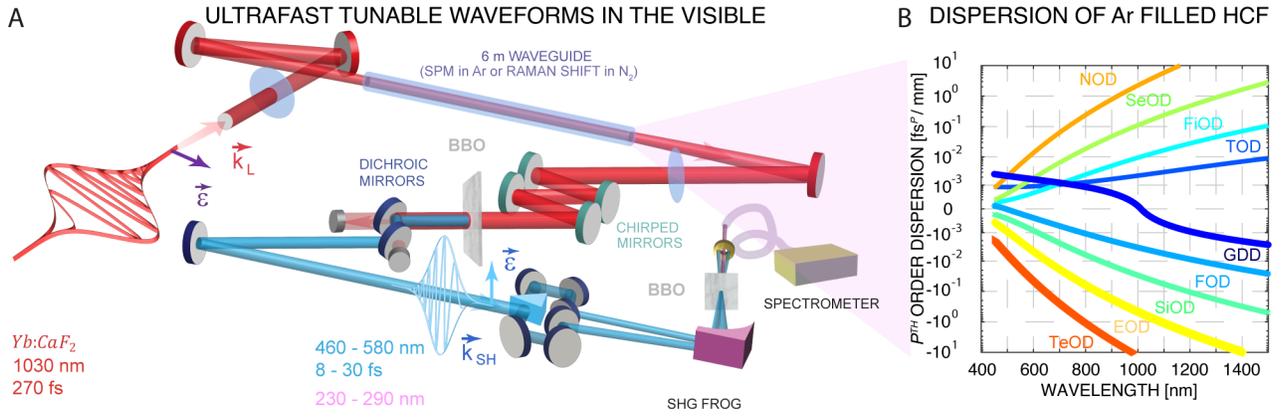

**Figure 1. Experimental setup for pulse compression and generation of tunable VIS pulses at 460 – 580 $nm$ or UV C pulses at 230 – 290 $nm$** A ) Sub-picosecond pulses at 1030 $nm$ are broadened and compressed to $15 - 30\ fs$ using a gas-filled hollow-core fiber and chirped mirrors. A $BBO$ crystal generates the VIS beams at 460 – 580 $nm$ with a duration of $8 - 30\ fs$. The pressure and the gas type in the waveguide control the dispersion of the broadened spectra. A second $BBO$ crystal is placed after the first one to upconvert the light to its second harmonic at shorter UV C wavelengths of 230 – 290 $nm$. B) Dispersion orders of the $Ar$ filled gas-waveguide system at $p = 80\ mbar$ pressure and room temperature, near the zero GDD dispersion for a 1030 $nm$ laser wavelength.

In this work, we demonstrate nonlinear self-compression by frequency upconversion to the second and fourth harmonic of chirped IR pulses resulting in a vastly wavelength-tunable ultrafast source in the visible and UV C spectral range and for the immediate goal of generating continuously wavelength-tunable light from the EUV to the soft-Xray spectral region. Moreover, our approach enables a high harmonic shift by two photons or one odd harmonic in the HHG comb in the EUV and soft-X-ray region, offering full tunability towards lower or higher energies while preserving the high HHG brightness. The details of the tunable X-ray source are demonstrated in a separate paper. [13].

**Wavelength-tunable, ultrafast waveforms in the VIS and UV C.** The development of high-energy, tunable ultrafast UV and VIS lasers has been largely constrained due to a deficiency of reliable optical modulators and fast coating degradation at shorter wavelengths. However, nonlinear optics have aided the realization of ultrashort pulses through frequency upconversion, direct harmonic generation, or parametric processes using IR ultrafast lasers at the cost of modest energy losses. In our experiments, we used an infrared $Yb:CaF_2$ laser amplifier at 1030 $nm$, delivering sub-picosecond laser pulses, $\sim 270\ fs$, with up to 14 $mJ$ at 0.5 $kHz$ (Fig. **1**) [14].

The pulses are first spectrally broadened in a 6-meter-long hollow-core fiber with a 1 $mm$ core diameter and compressed using a set of four chirped mirrors, balancing a group delay dispersion ($GDD$) of near 600 $fs^2$ [15-21]. Using such an approach, the $Yb:CaF_2$ laser pulses can be compressed below $15 fs$, with partially uncompensated higher orders. The laser wavelength, energy, pulse dispersion, fiber length, diameter and fiber photonic structure, gas density, and gas species influence the spectral broadening process. Fig. 2 shows more than 3000 fiber simulations that we used to optimize the spectral broadening in a wide parameter space (Luna) [22]. The asymmetry of the broadening due to the group delay dispersion (GDD) added to the driving pulse illustrates the amount of dispersion the system adds in the self-phase modulation (SPM) process, Fig. 2A.

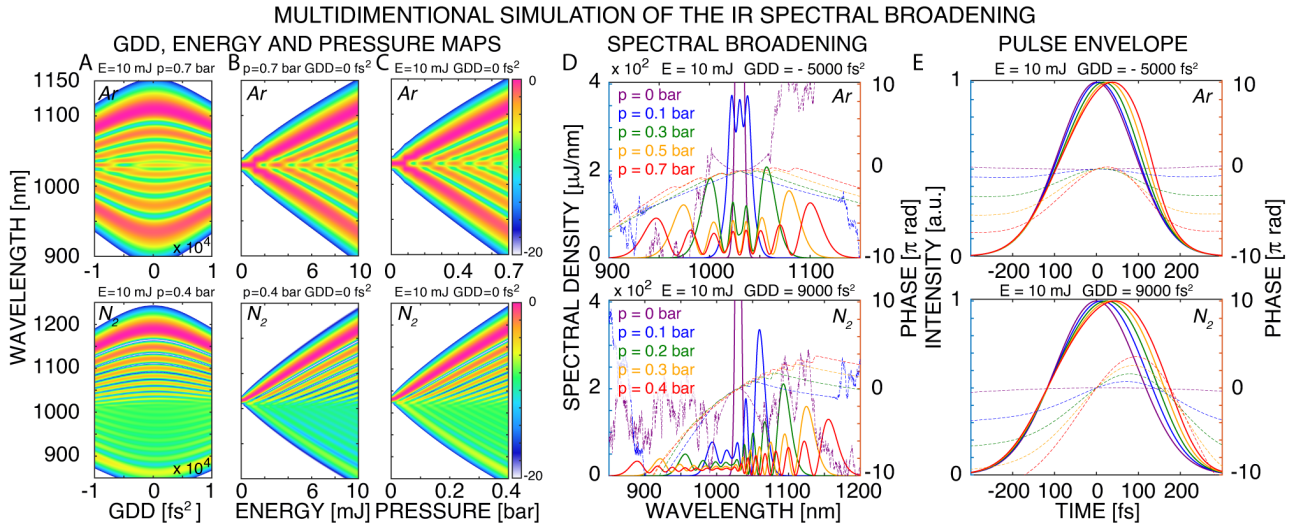

**Figure 2. Optimization of spectral broadening of a femtosecond 1030 $nm$ Yb-based laser using atomic and molecular gases.** A), B), and C) Two-dimensional logarithmic maps of spectral broadening showing symmetric behavior in atomic gas and asymmetric in molecular gas when varying either second-order dispersion GDD, pulse energy, or gas pressure in a 6-meter long hollow waveguide. The constant parameters are set to $E = 10\ mJ$, $GDD = 0\ fs^2$, $p = 0.7\ bar$ of $Ar$, and $p = 0.4\ bar$ of $N_2$. D) Spectra and phases at the waveguide output for specific pressures at $E = 10 mJ$, $GDD = -5000 fs^2$ (Ar) and $GDD = 9000 fs^2$ ($N_2$). E) Pulses and phases at the same conditions as in D)

The molecular nitrogen gas is less sensitive to variations in the GDD than the atomic Ar gas. Fig. 2 D shows the characteristic oscillations of the SPM-induced broadening for the atomic gasses and the Raman-enhanced asymmetric broadening towards the infrared.

The oscillations located at the farthest ends of the spectrum display a smooth and consistent phase, indicating the potential for naturally compressed pulses at specific pressure points within that range. However, the stepwise nature of the spectral phases poses a challenge for pulse compression as it leaves higher orders unbalanced when using basic compressors. We obtain tunable driving pulses in the visible spectral range of 460-580 nm with ultrashort pulse durations of $8 - 30\ fs$, using a $BBO$ type I second harmonic generation crystal with a conversion efficiency of $20 - 30\%$ (Fig. 3A).

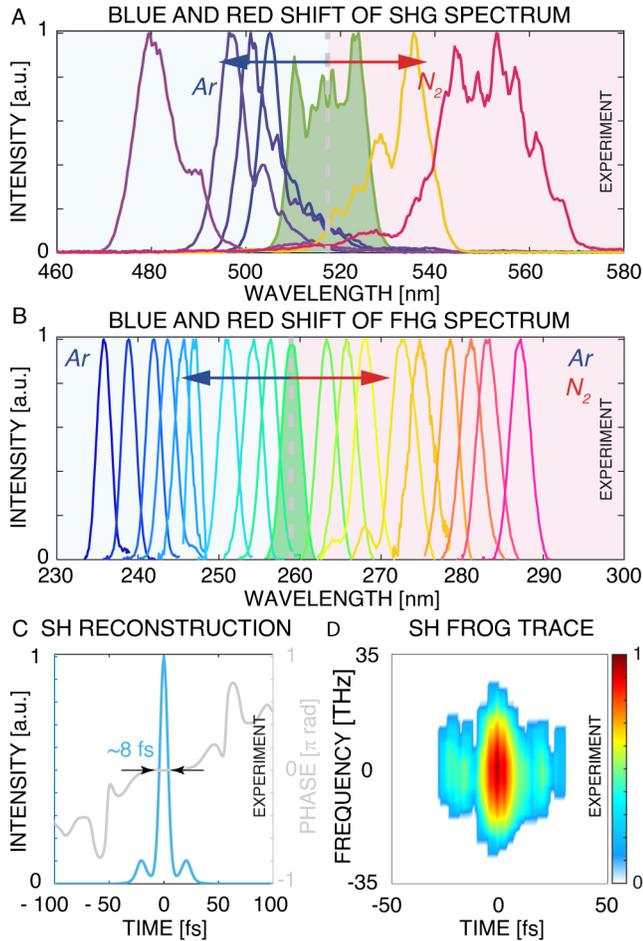

**Figure 3. Generation of ultrafast, tunable VIS pulses at 460 – 580 nm and UV C pulses at 230 – 290 $nm$.** A) Tunable second harmonic spectra with different central wavelengths in the range of 460 – 580 $nm$, optimized for high energy and short pulse durations of $8 - 30\ fs$ for a high-order harmonic generation. The pulses are generated in a $BBO$ crystal using $Ar$ (blue) or $N_2$ (red) in the waveguide with different pressures and $BBO$ angles. B) Tunable fourth harmonic spectra with different central wavelengths in the range of 230 – 290 $nm$ from spectrally broadened pulses in $Ar$ and $N_2$ C) Ptychographic pulse reconstruction of a record short duration of 8.4 $fs$. D) Measured FROG trace of the SHG pulse.

The BBO crystals are capable of phase-matching a finite bandwidth. In order to extract a second harmonic with the maximum energy and desired spectrum., we vary the gas species and adjust the pressure inside the fiber, energy, and laser compressor dispersion. The fiber is filled with either $Ar$ or $N_2$ gas at pressures ranging from $30 - 1000\ mbar$, providing extreme tunability of the blue or red-shifted spectral peak of the second harmonic beams.

Additionally, we extend the tunability of the source further into the UV C range of 230 – 290 nm by generating the fourth harmonic of the $Yb: CaF_2$ pulses, which produces pulses with an average transform-limited duration of 35 $fs$, Fig. 3B. Such a UV tunable source can enhance the free electron laser generation in space and time by providing a seed beam that matches the resonance condition of the electron bunches.

The tunable longer wavelength spectra in the second and fourth harmonic generation process are more favorably extended in molecular gasses. This is because most of the energy of the fundamental pulse is spectrally red-shifted in the asymmetric Raman broadening using molecular media. In contrast, the pulse energy is split between the red and blue branches in the symmetric SPM broadening using atomic media. In our experiments, atomic gasses such as $Ar, Kr$, and $Xe$ can be used to produce blue-shifted broadened spectra, while molecular gasses such as $N_2$ and air can be used to produce red-shifted broadened spectra. At the pressures of interest, the combined gas-waveguide system exhibits positive odd dispersion orders and negative even orders at the fundamental wavelength, except for the second order GDD, which switches sign from negative to positive at near $80\ mbar$ of $Ar$ or $N_2$, Fig. 1. In our pressure tuning range, a negative GDD contributes to soliton-like dynamics of the pulses, while a positive dispersion produces broad supercontinua, requiring external phase compensation for pulse compression. A near-zero dispersion results in pulse-splitting effects and additional frequency generation. We use a recently developed Lah-Laguerre analytical formalism to compute the first 10 orders of dispersion of our gas-waveguide system with the $p^{th}$ order chromatic dispersion evaluated as [23-25]:

$$POD = \frac{\partial^p}{\partial \omega^p} k(\omega) = \frac{(-1)^p}{c} \left(\frac{\lambda}{2\pi c}\right)^{p-1} \sum_{m=0}^{p} \mathcal{B}(p,m)\ \lambda^m \frac{\partial^m}{\partial \lambda^m} n(\lambda)$$

where $n(\lambda) = \sqrt{n_{gas}^2 - \left(\frac{\lambda u_{nm}}{2\pi a}\right)^2}$ is the effective refractive index of the system, $a$ is the radius of the fiber, $n_{gas}$ is the refractive index of the gas and $u_{nm} = 2.4048$ for the $HE_{11}$ mode. The matrix elements $\mathcal{B}(p,m)$ are the Laguerre coefficients of the order minus two.

Precise measurement of ultrashort pulses in the UV-visible spectral region can be challenging because any unnecessary pulse propagation in the air leads to pulse broadening. We use a compact second harmonic generation frequency-resolved optical gating (SHG FROG) placed near the SH harmonic generation to minimize pulse broadening. Furthermore, any accumulated phase can be subtracted from the measured signal to obtain a more accurate representation of the pulse duration [23, 24]. Fig. 3B and C show the measured and reconstructed pulse duration of a record short 8.4 fs pulse. The measurement was conducted using SHG FROG, and the VIS pulse was reconstructed with a ptychography algorithm [26-28].

To understand the origin of the short VIS pulse durations, we simulate the nonlinear pulse self-compression mechanism in a 300 $\mu m$ $BBO$ type I crystal and a chirped pulse with a broad spectrum that supports a 7 $fs$ transform-limited pulse at 1030 $nm$, with residual dispersion terms that are fit to the experimentally measured input and output pulse durations: $GDD$

of 90 $fs^2$, TOD of 500 $fs^3$ and fairly well-compensated values of the higher orders up to the tenth $(-1)^p\, 100\, fs^p$ with an alternating sign, where $p \geq 4$ is the dispersion order (Hussar) [21, 23-25, 29]. The transfer of energy between the fundamental and second harmonic pulses is constrained by the group velocity walk-off, along with the chromatic dispersion of the BBO material. This limitation causes the pulses to undergo temporal reshaping and spread, as depicted in Figure 4 (A). However, the situation changes when the pulse tail contains substantial energy. By using pulses with high order chirp, we show that the power from the rear of the fundamental pulse can be transferred to the second harmonic peak at larger propagation distances, owing to the group-velocity mismatch.

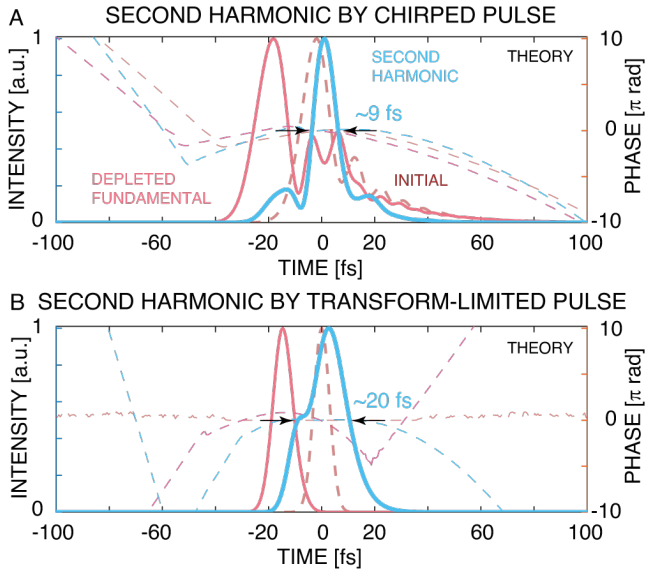

**Figure 4. Nonlinear pulse compression effect in second harmonic generation using 3+1D spatio-temporal beam propagation for pulses with high order dispersion.** A) Simulation of nonlinear pulse compression in the second-harmonic generation to 9 $fs$ at 515 $nm$ in a 300 $\mu m$ BBO type I crystal for a 1030 $nm$ input pulse of 7 $fs$ with the high order dispersion in Table 1. B) In contrast, a transform-limited input pulse of 7 $fs$ gets stretched to 20 $fs$ due to group velocity walk-off. The dashed red and solid red-rose lines show the input and output 1030 $nm$ pulses, respectively, and the blue lines show the second harmonic pulses. The dashed lines of matching colors illustrate the phases of the pulses.

This generates a sharp pulse peak with wings, however, with a shortened high-intensity part of the second harmonic pulse, containing most of the energy. The high-order dispersions of the input pulse combine uncompensated dispersion after the SPM or Raman spectral broadening, as well as high-order dispersion of the chirped mirrors. Hence, the $TOD$ and negligible $FOD$ and higher dispersion orders serve as driving sources for self-compression. Furthermore, the presence of high-order chirp prevents further strong distortion of the pulse during propagation in dispersive media. In contrast, a transform-limited fundamental pulse would require an extremely short crystal length to maintain a short pulse duration at the second harmonic, leading to low conversion efficiency.

A 300 $\mu m$ $BBO$ crystal is better suited for transform-limited pulses with around 20 $fs$ duration, producing around 18 $fs$ second harmonic pulses.

**Table 1. Numerically extracted dispersion orders from the simulations.**

| GDD | TOD | FOD | FiOD | SiOD | SeOD | EOD | TeOD | TeOD |
| [$fs^2$] | [$fs^3$] | [$fs^4$] | [$fs^5$] | [$fs^6$] | [$fs^7$] | [$fs^8$] | [$fs^{10}$] | [$fs^{10}$] |
| 28.7 | 3.3E3 | -1.1E4 | 2.9E04 | -8E04 | 2.3E5 | -7E+5 | 2.3E6 | -7.9E6 |

The transform-limited 7 $fs$ pulse with the spectrum considered in Fig. 4A, will generate a second harmonic with 20 fs pulse duration and a significant leading and trailing edge reshaping because of the group velocity walk-off and the inevitable fast pulse separation (Fig 4 (B)). This scenario leads to more energy transfer to the second harmonic pulse sides. The generation of a shortened second harmonic pulse in our experiments using a chirped fundamental pulse is accompanied by a very slight decrease in conversion efficiency. In outlook, the favorable interplay of using combinations of optimized pressure in the waveguide and phase-matching angle of the SHG crystal to tune the central UV-VIS laser wavelength over a vast range makes it possible to generate high harmonics with maximum energy conversion and continuously tunable comb peaks with an either blue or red spectral shift of interest for resonant X-ray applications.

Interestingly, intense EUV, UV, and visible lasers are ideal for harmonic generation since they combine favorable single-atom quantum physics and macroscopic phase-matching physics. High harmonics from short-wavelength lasers have many advantages. For example, very high efficiency, ultra-narrow bandwidths with large energy separation, effective phase and group velocity matching, excellent spatio-temporal coherence, and most importantly, extended X-ray cutoff with natural attosecond pulse structure [7, 30].

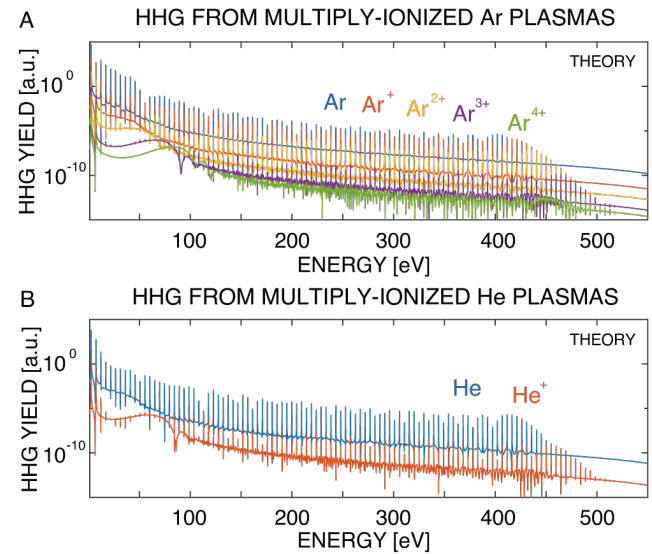

**Figure 5. Single-atom HHG yield from multiply ionized $Ar$ and $He$ plasmas generated by a VIS driving laser.** Strong field approximation estimate for the high harmonic yield in the soft X-ray regime driven by a VIS 515 $nm$ laser with moderate intensity of $5 \times 10^{15}\ W/cm^2$ and 8 $fs$ pulse duration in A) $Ar$ and B) $He$.

Using modest laser parameters of 1 $mJ$, $M^2 = 10$ and 8 $fs$ pulse duration at 515 $nm$, we theoretically illustrate that narrow-bandwidth harmonics >500 $eV$ can be produced from multiply ionized plasmas in $Ar$ (Fig. 5 (A)) and He (Fig. 5 (B)), while macroscopic phase matching will limit the usable highest photon energies to lower cutoffs.

Raytracing reveals space-time focusing is possible up to $f = 10\ cm$ focal length, producing the intensity of 5x10$^{15}$ W/cm2 used in these calculations [13, 31]. Since short-pulse high-energy optical parametric amplifiers are not available in the UV-VIS, the efficiency of a wavelength-tunable X-ray source using short-wavelength drivers can be straightforwardly investigated using the demonstrated approach. Our analysis indicates that the X-ray pulses appear as a train of relatively short 190 $as$ bursts with low attosecond chirp [7, 12]. Finally, these tunable UV-VIS sources are of interest for attosecond electron correlation spectroscopy using high harmonic generation. Recent experimental and theoretical studies demonstrated that using 400 nm drivers enhances the probability of observing attosecond electron-electron correlation effects in high-order harmonic generation spectroscopically in He atoms.

**Conclusion.** In summary, we demonstrate continuously wavelength-tunable UV and VIS pulses with ultrashort pulse duration as short as 8.4 fs through direct nonlinear self-compression. The immediate goal of our work is the generation of bright, narrowband, continuously wavelength-tunable harmonics, which can resonantly cover the magnetic N and M absorption edges of ferromagnetic and non-magnetic elements in the EUV range up to 100 eV. In addition, tunable harmonics in the 100 – 200 eV soft X-ray range have immense potential for studying the dynamics in advanced materials with sub-10 nm spatial resolution, as well as investigating attosecond electron correlations in strongly correlated systems, all on a tabletop scale. Moreover, these harmonics emerge as sub-200 attosecond pulse trains in the time domain, as indicated by calculations, allowing for dynamic pump-probe multidimensional (4+1D) imaging in the near future, where the additional effective dimension is the material identity. Finally, such UV or VIS tunable sources can directly seed free electron lasers and enhance high-intensity coherent radiation in the X-ray range.

**Funding.** H2020 European Research Council (100010663), XSTREAM-716950, Alfred P. Sloan Foundation (100000879) FG-2018-10892

**Acknowledgments.** TP acknowledges funding from the European Research Council (ERC) under the European Union's Horizon 2020 research and innovation program (grant agreement XSTREAM-716950), and from the Alfred P. Sloan Foundation (FG-2018-10892).

**Disclosures.** The authors declare no conflict of interest.

**Data availability.** Data underlying the results are presented in the plotted graphs.